\documentstyle[12pt,aasms4,epsfig,graphics,times,mathptmx]{article}

\def\gsim{\;\rlap{\lower 2.5pt
 \hbox{$\sim$}}\raise 1.5pt\hbox{$>$}\;}
\def\lsim{\;\rlap{\lower 2.5pt
   \hbox{$\sim$}}\raise 1.5pt\hbox{$<$}\;}

\def\ge{\;\rlap{\lower 2.5pt
 \hbox{$-$}}\raise 1.5pt\hbox{$>$}\;}
\def\le{\;\rlap{\lower 2.5pt
   \hbox{$-$}}\raise 1.5pt\hbox{$<$}\;}

\newcommand\beq{\begin{equation}}
\newcommand\eeq{\end{equation}}

\def\v{\vspace{-0.1in}}

\begin{document}

\Large \centerline{\bf Energy input from quasars regulates}
\centerline{\bf the growth and activity of black holes and their host galaxies}


\normalsize
\author{\bf Tiziana Di Matteo$^\dagger$\footnote{Present Address: 
Carnegie-Mellon University, Dept. of Physics, 
5000 Forbes Ave., Pittsburgh, PA 15213}, Volker Springel$^\dagger$ and
  Lars Hernquist$^\star$}
\medskip
\noindent
$\dagger$ Max-Planck-Institut f{\" u}r Astrophysik, Karl-Schwarzschild-Str.~1, 85740 Garching bei M{\" u}nchen, Germany \\ 
$\star$ Astronomy Dept., Harvard University, 60 Garden
Street, Cambridge, MA 02138, USA\\

\vskip 0.2in 
\hrule 
\vskip 0.2in

{\bf In the early Universe, while galaxies were still forming, black
  holes as massive as a billion solar masses powered quasars.
  Supermassive black holes are found at the centers of most galaxies
  today\cite{KR95}$^,$\cite{M98}$^,$\cite{FH04}, where their masses
  are related to the velocity dispersions of stars in their host
  galaxies and hence to the mass of the central
  bulge of the galaxy\cite{FM00}$^,$\cite{G00}. This suggests a link
  between the growth of the black holes and the host galaxies
  \cite{KH00}$^,$\cite{VHM03}$^,$\cite{Wy03}$^,$\cite{Gra04},
  which has indeed been assumed for a number of years.
  But the origin of the observed
  relation between black hole mass and stellar velocity dispersion,
  and its connection with the evolution of galaxies have
  remained unclear. Here we report hydrodynamical simulations that
  simultaneously follow star formation and the growth of black holes
  during galaxy-galaxy collisions. We find that
  in addition to generating a burst of star formation\cite{MH96}, 
  a merger leads to strong inflows that feed gas to the
  supermassive black hole and thereby power the quasar. The energy released by
  the quasar expels enough gas to quench both star formation and
  further black hole growth. This determines the lifetime of the
  quasar phase (approaching 100 million years) and explains the relationship
  between the black hole mass and the stellar velocity dispersion.}

A large fraction of the black hole mass in galaxies today is thought
to have been assembled during the peak of quasar activity in the early
Universe\cite{S82}$^,$\cite{YT02}, when large amounts of matter were
available for accretion onto central black holes.  Interactions and
mergers between galaxies are known to trigger large-scale nuclear gas
inflows\cite{H89}$^,$\cite{BH92}, which is a
prerequisite for the growth of central black holes by accretion. 
Also, hierarchical models of galaxy formation imply that
mergers of galaxies form elliptical galaxies or spheroidal components in
galaxies, by destroying stellar disks and triggering nuclear starbursts.

This led to suggestions that the $M_{\rm BH} - \sigma$ relation
(where $M_{\rm BH}$ is the black hole mass and $\sigma$ is the
velocity dispersion of stars in the bulge of galaxies) could
arise in galaxy mergers, provided strong outflows are produced in
response to major phases of accretion, capable of halting further
black hole
growth\cite{SR98}$^,$\cite{F99}$^,$\cite{K03}$^,$\cite{Wy03}.  Indeed,
observations point to the existence of strong outflows in bright
quasars\cite{CBG03}$^,$\cite{CKG03}$^,$\cite{P03}. In this picture,
black hole accretion is expected to have a crucial impact on the
evolution of the host galaxy. However, the coupling of star formation
with black hole growth in the context of galaxy evolution is difficult
to treat on the basis of analytical estimates alone.

We have, therefore, performed detailed numerical simulations of galaxy
mergers that include radiative cooling, star formation, black hole
growth, energetic feedback from supernovae and accretion onto black
holes, as well as the gravitational dynamics of gas, stars, and dark
matter (see Supplementary Information and further details in
ref.~21). As described in the supplementary information, we
include a novel treatment of gas accretion onto supermassive black
holes and its associated feedback in the centres of merging galaxies.
We describe black holes using collisionless ``sink'' particles that
can grow in mass by accreting gas from their surroundings. The
accretion rate is estimated by relating the small-scale, unresolved
flow around the black hole to the large-scale, resolved gas properties
using a spherical Bondi-Hoyle\cite{B52} model.  We further assume that
a small fraction $f$ of the radiated luminosity $L$ couples
thermodynamically to the surrounding gas. This gives an effective
heating rate, $ \dot{E}_{\rm feed} = f L = f \eta \dot{M} c^2$, where
we fix the radiative efficiency to $\eta = 0.1$, the typical value
derived from Shakura-Sunyaev\cite{SS73} accretion models onto a
non-rotating black hole. We further assume $f=0.05$, so that $\sim 0.5
\%$ of the accreted rest mass energy is available to heat the gas.

To illustrate the impact of central, supermassive black holes on
mergers of two disk galaxies, we compare the gas evolution between two
simulations with galaxies which have roughly the size of the Milky Way
(Fig.~\ref{fig1}). Star formation and supernova feedback are included
in both simulations, but we add our black hole growth and feedback in
one (top four panels of Fig.~\ref{fig1}, see also the corresponding
Supplementary Video) and neglect it in the other (bottom four
panels). The first image ($t=1.1$ Gyr) shows the galaxies soon after
their first encounter, when strong gravitational forces have spawned
extended tidal tails. At this time, the black holes in the centres of
each galaxy (top) have already grown significantly from their initial
masses and are accreting at a moderate level. However, the overall
star formation rate is essentially unaffected by the presence of the
black holes, as indicated by Figure~\ref{fig2}, which plots the star
formation rate (in both cases), black hole accretion rate, and black
hole mass (for the top panels), as a function of time.

The second snapshot ($t=1.4$ Gyr) in Fig.~\ref{fig1} shows the
galaxies when they begin to coalesce. Here, the tidal interaction has
distorted the disks into a pair of bi-symmetric spirals, and gas is
shocked between the two galaxies. The tidal response drives gas into
the central regions of each galaxy. While only weak starbursts and
accretion events are triggered at this time, it is already evident
that black hole feedback alters the thermodynamic state of the gas, as
indicated by the relatively lower density and higher temperature of
the gas surrounding the galaxies in the simulation with black holes.
In fact, a significant wind has started to flow out of the centres.

When the galaxies finally merge, as shown in the third image ($t=1.6$
Gyr), much of the gas is quickly converted into stars in intense
bursts of star formation\cite{MH96}. Owing to the enhanced gas
density, the black holes, which also merge to form one object,
experience a rapid phase of accretion close to the Eddington rate
resulting in significant mass growth. Also, the morphologies of the
remnants in the two simulations begin to differ significantly. In the
model without black holes, most of the gas is still inflowing in a
comparatively cool phase. In contrast, the simulations with black
holes exhibit a significant change in the thermodynamic state of the
circumnuclear gas, which is heated by the feedback energy provided by
the accretion and partly expelled in a powerful wind. During this
strong accretion phase and for this interval of time, the object would
be a bright quasar with a specific lifetime.

Differences persist as the remnants settle into a relaxed state
(fourth panels in Fig.~\ref{fig1}). The remnant without black holes
retains a large amount of dense cold gas, yielding prolonged star
formation at a steady rate. However, in the simulation with
supermassive black holes, nearly all the gas is expelled from the
centre, quenching star formation and black hole accretion itself.
Consequently, the black hole mass saturates, quasar activity stops,
and star formation is inhibited, so that the remnant resembles a
``dead'' elliptical galaxy whose stellar population quickly
reddens\cite{SDH04}.  In the particular example we show, the remnant
of this major merger is an elliptical galaxy. In the hierarchical
model of galaxy formation, a new disk can grow around this spheroid,
turning it into the bulge component of a spiral galaxy.  In very
gas-rich mergers, a disk component may even survive
directly\cite{SH04} (an example of this is shown in a Supplementary
Figure). We therefore expect black holes in bulges of spiral galaxies
to be assembled in a manner similar to those in ellipticals. Our
results should also apply for cases involving minor mergers.

The evolution of star formation rate, black hole accretion rate, and
black hole mass for mergers when the progenitor galaxy mass is varied
(including the model shown in Fig.~\ref{fig1}) are shown in
Figure~\ref{fig2}. Models with different mass qualitatively reproduce
the key features of the evolution shown in Figure~\ref{fig1}: the star
formation and black hole accretion rates are both quenched in the
remnant, and black hole growth saturates owing to feedback provided by
accretion energy.  However, the damping of star formation and black
hole activity is more abrupt in the more massive systems. Here, the
total gas supply for accretion is larger, and the gravitational
potential well is deeper, and so the black hole has to grow much more
before its released energy is sufficient to expel the gas in a quasar
driven wind, which then terminates further nuclear accretion and star
formation. For the same reasons, the initial growth of the black
holes, which is regulated by the properties of nearby gas, depends on
the total mass. It is faster in more massive systems, which can
therefore reach the exponential, Eddington-limited growth phase more
easily. The lifetime of the active black hole phase, however,
increases for smaller black hole masses, implying that low-luminosity
quasars should be more numerous than bright ones. This is consistent
with them residing in smaller galaxies and with what has been found in recent
surveys\cite{HMS04}$^,$\cite{Bar04}.

The dependence of black hole growth on galaxy mass yields a relation
between the stellar spheroid of the remnant and its central black
hole. Figure~\ref{fig3} shows the black hole mass versus the stellar
velocity dispersion of the merger remnants from our simulations,
compared with observations. We show simulations with six different
galaxy masses, each of which has been run with three different initial
gas mass fractions of the galaxies' disks.  Remarkably, our simulations
reproduce the observed $M_{\rm BH}$--$\sigma$ correlation very well.
Note that black holes in more gas-rich mergers reach somewhat larger
masses than those growing in gas-poorer environments (which is
expected from our prescription for the accretion rate), but this is
partly compensated by an increase in the velocity dispersion of the
corresponding bulges, maintaining a comparatively tight $M_{\rm
  BH}$--$\sigma$ relation. However, this suggests that part of the
intrinsic scatter in the observed relation can be ascribed to
different gas fractions of the galaxies during black hole growth. Our
lowest mass galaxy models probe a region of the $M_{\rm BH}$--$\sigma$
relation where few measurements are available and predict that this
correlation should hold towards small black hole masses and velocity
dispersions, in tentative agreement with recent
observations\cite{GH04}.

Black hole growth is self-regulated in our models. As galaxies merge
to form spheroids, the dynamical response of the gas to the energy
supplied by accretion halts further growth once the black holes have
reached a critical size for the gravitational potential of the bulge.
At this saturation point, the active galactic nuclei (AGN) generate
outflows that drive away gas and inhibit further star formation. Our
simulations are the first self-consistent models to demonstrate that
self-regulation can quantitatively account for the principle
observational facts known for the local population of supermassive
black holes, most notably the $M_{\rm BH}$--$\sigma$ relation.
Moreover, self-regulation in our hydrodynamical simulations predicts a
specific duration of the luminous episode of a black hole in a given
galaxy, thereby explaining the origin of quasar lifetimes.  We note
that the final black hole masses we obtain are roughly proportional to
the inverse of the value assumed for the feedback efficiency, $f$.
Interestingly, our choice of $f =5 \%$ is consistent with the value
required in semi-analytic models\cite{Wy03} to explain the evolution
of the number density of quasars.

The black hole accretion activity also has a profound impact on the
host galaxy. The remnant spheroid is gas-poor and has low residual
star formation, so it evolves to a red stellar colour on a short
timescale. The simulations shown here make it possible to draw firm
conclusions on this and other links between black hole growth, quasar
activity, and properties of the galaxy population. Our novel approach
can also be implemented in cosmological simulations of hierarchical
structure formation in representative pieces of the Universe.  Such
simulations will allow us to study directly why quasars were
much more numerous in the early Universe than they are today, and how
black holes and galaxies influence each other throughout cosmic
history.

\vskip 0.2in
\noindent
{\bf Supplementary Information} accompanies the paper on 
{\bf www.nature.com/nature}

\normalsize
\vskip 0.2in
\noindent
{ACKNOWLEDGEMENTS.} 
The computations reported here were performed at the 
Center for Parallel Astrophysical Computing at
Harvard-Smithsonian Center for Astrophysics and at the 
Rechenzentrum der Max-Planck-Gesellschaft in Garching.

\vskip 0.2in
\noindent Send correspondence and requests for materials to T.D.M.
  (email: tiziana@phys.cmu.edu)

\vskip 1in

\clearpage

\begin{figure*}[htbp]
\caption{\label{fig1} 
  Snapshots of the simulated time evolution
  in mergers of two galaxies with and without black holes. The model
  with black holes is shown in the top panels. The full time sequence
  for this simulation can be viewed in the Supplementary Video. The
  bottom row shows the corresponding simulation without the inclusion
  of black holes. In both cases, four snapshots at different times in
  the simulations are shown. The images visualise the projected gas
  distribution in the two galaxies, colour-coded by temperature (blue
  to red).  The colliding galaxies have the same initial mass
  corresponding to a `virial velocity' of $V_{\rm vir} = (M_{\rm tot}
  \times 10 G H_{0} )^{1/3} = 160$ km/s, and consist of an extended
  dark matter halo, a stellar bulge, and a disk made up of stars and
  20\% gas.  Each individual galaxy in the simulations is represented
  with 30000 particles for the dark matter, 20000 for the stellar
  disk, 20000 for the gaseous disk, and 10000 for the bulge component.
  Two such galaxies were set up on a parabolic, prograde collision
  course, and then evolved forward in time numerically with {\small
  GADGET-2}\cite{SH03}, a parallel TreeSPH simulation code.  The first
  snapshot ($t=1.1$ Gyr) shows the systems after the first passage of
  the two galaxies. The second snapshot ($t=1.4$ Gyr) depicts the
  galaxies distorted by their mutual tidal interaction, just before
  they merge.  The peak in the star formation and black hole accretion
  (see also Fig.~2) is reached at the time of the third snapshot
  ($t=1.6$ Gyr), when the galaxies finally coalesce. At this time, a
  strong wind driven by feedback energy from the accretion expels much
  of the gas from the inner regions in the simulation with black
  holes. Finally, the last snapshots show the systems after the
  galaxies have merged ($t=2.5$ Gyr), leaving behind quasi-static
  spheroidal galaxies. In the simulation with black holes, the remnant
  is very gas poor and has little gas left dense enough to support
  ongoing star formation. This highlights that the presence of
  supermassive black holes, which accrete from the surrounding gas and
  heat it with the associated feedback energy, dramatically alters the
  merger remnant.}
\end{figure*}

\clearpage

\begin{figure*}[htbp]
\begin{center}
\resizebox{10.0cm}{!}{\includegraphics[angle=270]{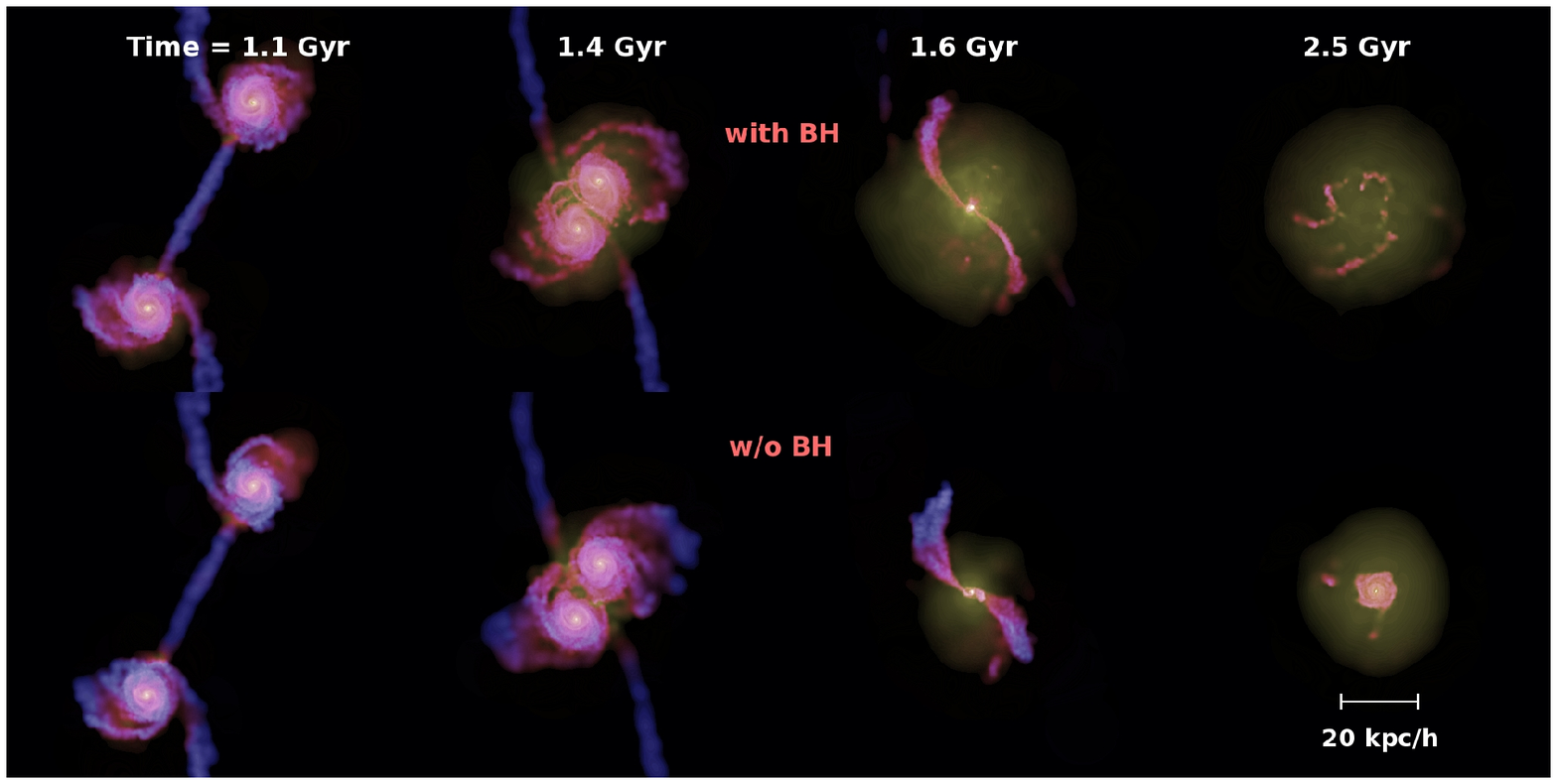}}
\end{center}
\end{figure*}

\clearpage

\begin{figure*}[htbp]
\caption{\label{fig2} 
  Black hole activity, star formation and black hole growth plotted as
  a function of time during a galaxy-galaxy merger. The star
  formation rate (SFR) and black hole accretion rate (BHAR) are shown
  in the top and middle panels, respectively, and are given in units
  of solar masses per year. The black hole mass ($M_{\rm BH}$). is
  given in units of solar masses. The three lines in each panel
  correspond to models with galaxies of virial velocity $V_{\rm vir} =
  80$, $160$, and $320$ km/s, (bottom to top lines in each panel, also
  labelled in the bottom panel). For comparison, we also show the
  evolution of the star formation rate for the model without a black
  hole that is shown in Fig.~\ref{fig1} (dashed line -- for the
  $V_{\rm vir} = 160$ km/s galaxy). We note, in particular, that owing
  to AGN feedback, the peak amplitude of the starburst during the
  merger is lowered by a significant factor. The black solid circles
  in the individual panels identify the times of the corresponding
  snapshots shown in Figure~1.}
\end{figure*}

\clearpage

\begin{figure*}[htbp]
\resizebox{18.0cm}{!}{\hspace{-2cm}\includegraphics{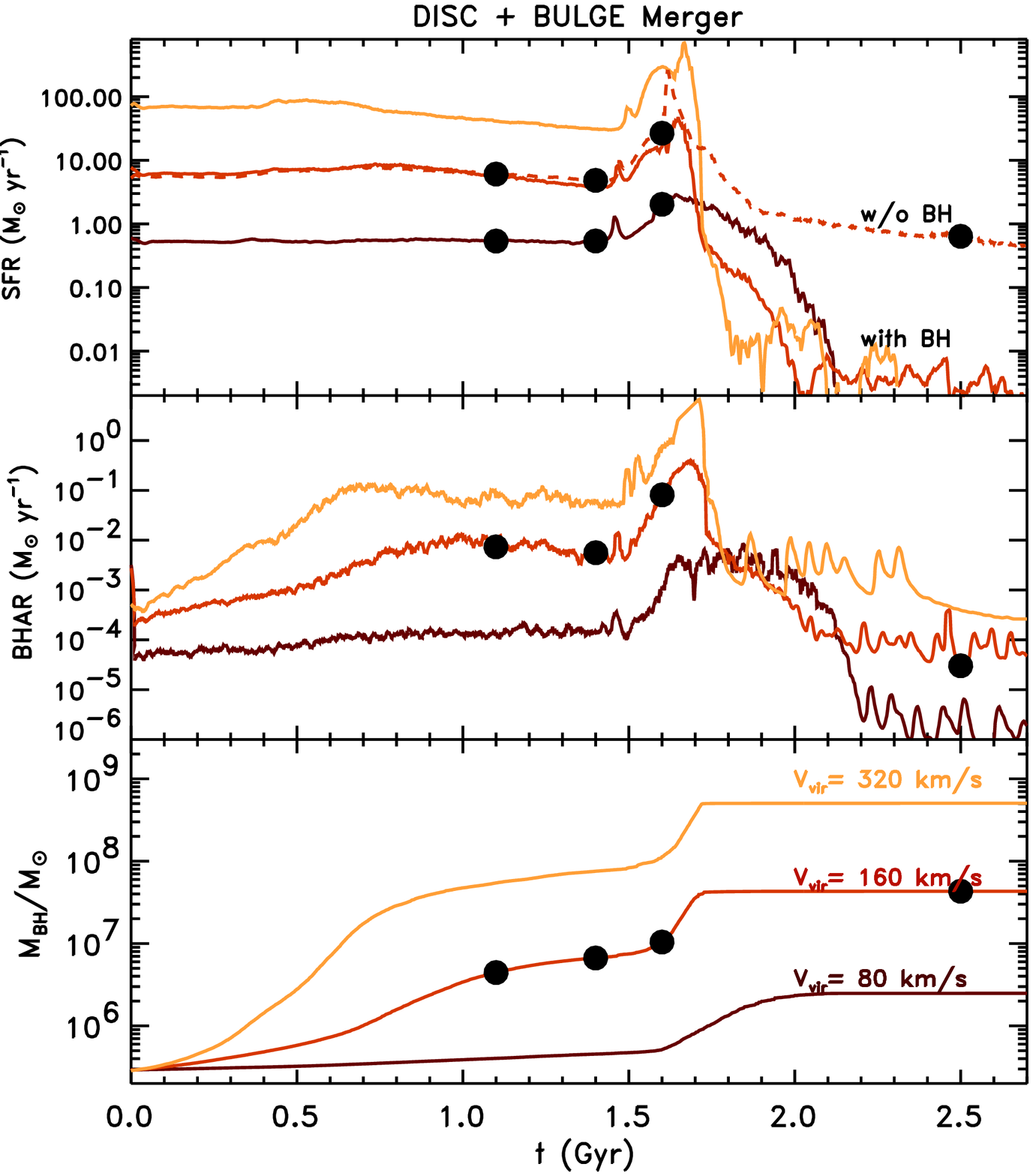}}
\end{figure*}

\clearpage

\begin{figure*}[htbp]
\caption{\label{fig3} The relation between the final black hole
  mass, $M_{\rm BH}$, and the velocity dispersion of stars, $\sigma$,
  of our galaxy merger simulations compared with observational
  measurements.  The solid circles show the masses of the black holes
  and the bulge velocity dispersions measured for the final remnants
  of six merger simulations of galaxies with disk gas fraction of
  20\%, but different total mass, parameterised by virial velocities
  of $V_{\rm vir} = 50$, $80$, $160$, $320$, and $500\;{\rm km/s}$ (shown
 by the dark to light red, from low to high mass galaxies respectively).
  Open circles and open squares with the same colour give results for
  gas fractions of 40\% and 80\%, respectively. We have also checked
  that our results are insensitive to the orbits of the galaxy
  collisions.  Mimicking the observational data, we calculate $\sigma$
  as the line-of-sight stellar velocity dispersion of stars in the
  bulge within the effective radius, $R_{e}$, of the galaxy.  Black
  symbols show observational data for the masses of supermassive black
  holes and the velocity dispersions of their host bulges.
  Measurements based on stellar kinematics are denoted by filled
  stars, those on gas kinematics by open squares, and those on maser
  kinematics by filled triangles. Details for all the displayed
  measurements are given in ref.~{3} and {28}. The observed BH sample
  has been fit by a power law relation, yielding\cite{T02}: $M_{\rm
    BH}=(1.5 \pm 0.2) \times 10^{8} {\rm M}_{\odot} ({\sigma}/{200\,
    {\rm km/s} })^{4.02 \pm 0.32}$.  The inset shows the relation
  between the circular velocity $V_{\rm vir}$ and $\sigma$ measured
  for the merger remnants in the simulations. The same colour coding is
used as in the main panel to indicate corresponding mass objects.}
\end{figure*}

\clearpage

\begin{figure*}[htbp]
\resizebox{18.0cm}{!}{\hspace{-2cm}\includegraphics{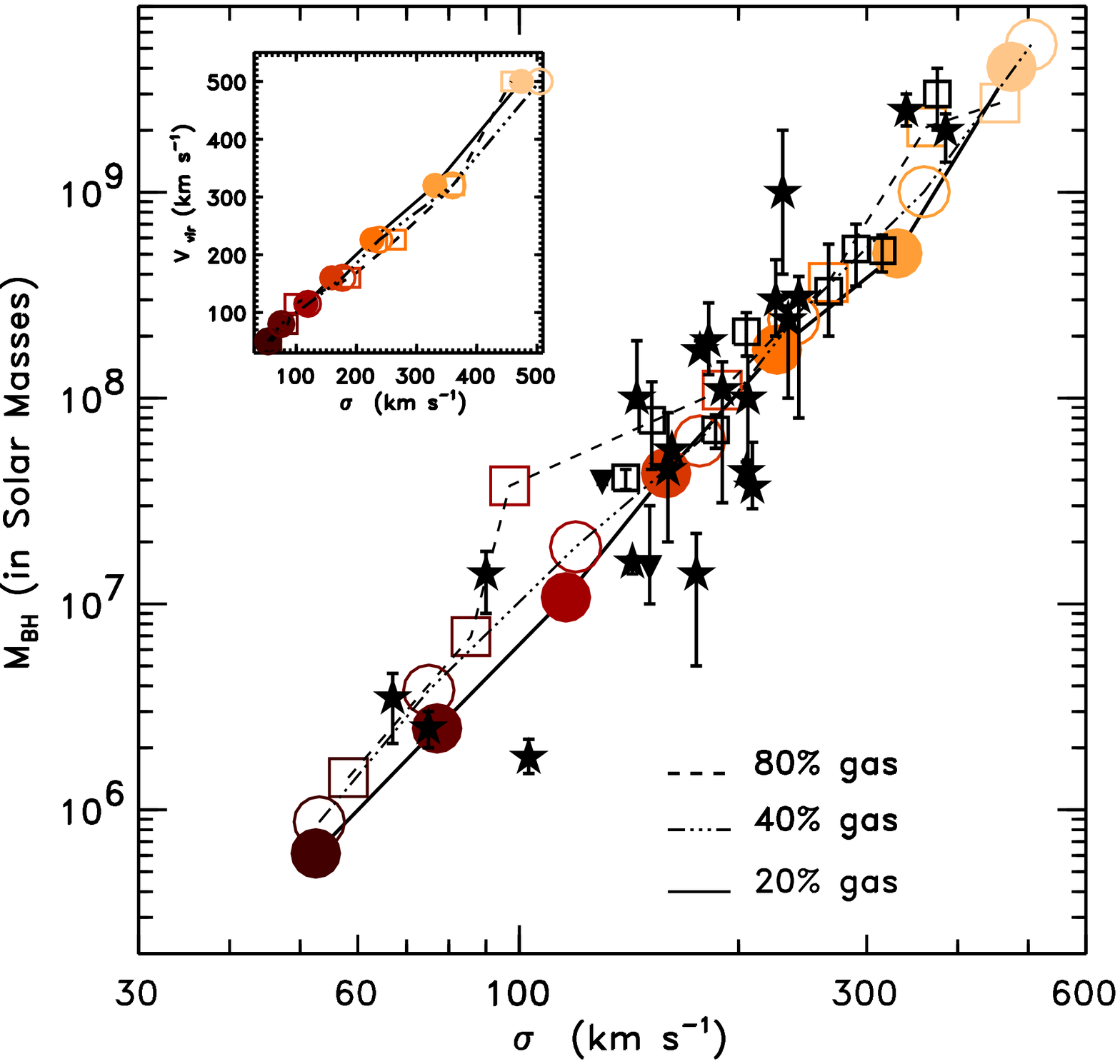}}
\end{figure*}

\end{document}